\documentclass{Interspeech2024}
\interspeechcameraready

\usepackage{multirow}
\usepackage{amsmath}
\usepackage{amssymb}
\usepackage{dsfont}
\usepackage{booktabs}
\usepackage{multicol}
\usepackage{multirow}

\title{Guiding Frame-Level CTC Alignments Using Self-knowledge Distillation}

\name[affiliation={1*}]{Eungbeom}{Kim}
\name[affiliation={2\dagger}]{Hantae}{Kim}
\name[affiliation={1,3}]{Kyogu}{Lee}

\address{
  $^1$IPAI, Seoul National University, Republic of Korea\thanks{* Work done during internship at NAVER Cloud.}\\
  $^2$NAVER Cloud, Republic of Korea\thanks{$^\dagger$ Corresponding author.} \\
  $^3$Dept. of Intelligence and Information, AIIS, Seoul National University, Republic of Korea}
\email{eb.kim@snu.ac.kr, hantae.kim@navercorp.com, kglee@snu.ac.kr}


\begin{document}

\maketitle

\begin{abstract}
    Transformer encoder with connectionist temporal classification (CTC) framework is widely used for automatic speech recognition (ASR). However, knowledge distillation (KD) for ASR displays a problem of disagreement between teacher-student models in frame-level alignment which ultimately hinders it from improving the student model’s performance.
    In order to resolve this problem, this paper introduces a self-knowledge distillation (SKD) method that guides the frame-level alignment during the training time. In contrast to the conventional method using separate teacher and student models, this study introduces a simple and effective method sharing encoder layers and applying the sub-model as the student model.
    Overall, our approach is effective in improving both the resource efficiency as well as performance. We also conducted an experimental analysis of the spike timings to illustrate that the proposed method improves performance by reducing the alignment disagreement.


    

\end{abstract}

\section{Introduction}
Automatic speech recognition (ASR) is a sequence labeling task that aims to translate a given speech utterance into a transcript~\cite{graves2012sequence, hannun2014deep, chan2016listen, rao2017exploring, vaswani2017attention}. To train the ASR model, Connectionist Temporal Classification (CTC)~\cite{graves2006connectionist} is a widely used method when the speech-transcript alignment is unknown, which is a common setting for real-world data. By combining CTC and scalable self-supervised learning methods, recent studies \cite{schneider2019wav2vec, baevski2020wav2vec, hsu2021hubert, babu22_interspeech,  baevski2022data2vec, chen2022wavlm} lead to a substantial performance gain with large-scale ASR architectures.

Moreover, the large-scale models are computationally expensive and memory-inefficient. For this reason, various methods including knowledge distillation (KD)~\cite{hinton2015distilling} are applied to attain competitive results under limited environment. Conventional KD leverages a large teacher model to train a small student model, in which the student and the teacher models have independent models. 
In addition, \cite{zhang2019your, xu2019data, ji2021refine} explain the self-knowledge distillation framework (SKD) which distills the knowledge of the model within itself.

Although CTC enables us to train the ASR models successfully without the frame-level input-label alignment, the alignment estimated by CTC can be unstable and arbitrary~\cite{sak2015learning, kurata2019guiding, tian2023bayes}. Thus, directly applying the knowledge distillation of the teacher model's frame-level alignment to the student model may possess two major risks in performance degradation~\cite{sak2015learning}.

The first issue is originated from instability of the trained alignment. Since a single transcript can be mapped over various alignments, the teacher and student models' alignments can also be diverse. This alignment disagreement distracts the knowledge distillation process which disturbs the convergence of the student model~\cite{takashima2018investigation, yoon2021tutornet}. To mitigate this issue, several methods are proposed. \cite{takashima2018investigation, takashima2019investigation} only utilize sequence-level information extracted from the frame-level alignment. However, the removal of useful frame-level information results in a limited performance increase. Although Guide-CTC~\cite{kurata2019guiding} distills the useful frame-level information from the teacher model alignment, teacher-student alignment disagreement remains problematic~\cite{tian2022knowledge}.

Likewise, most of the previous methods are limited to post-processing of the teacher model's alignment because the previous methods adhere to the teacher-student knowledge distillation framework from which the alignment disagreement problem stems. This study aims to alleviate this alignment disagreement problem in advance by proposing an in-processing solution. In order to achieve our goal, we explore a new knowledge distillation strategy for CTC based on a self-knowledge distillation upon the intermediate CTC~\cite{lee2021intermediate} method. Since self-knowledge distillation method such as~\cite{zhang2019your} includes teacher and student in the same model, with the teacher inheriting the student's output sequences, the disagreement in the teacher-student alignments is intrinsically mitigated compared to the independent teacher and student framework. We experimentally verify this argument by quantitatively comparing the alignments of the teacher and student models for the self-knowledge distillation and the conventional knowledge distillation.

\begin{figure*}[h!]
    \centering
    \includegraphics[width=\textwidth]{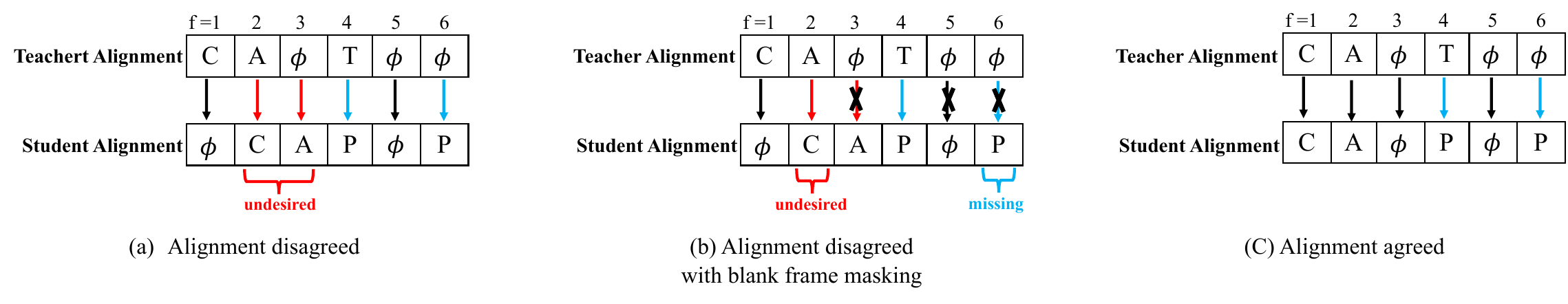}
    \caption{Illustrations of knowledge distillation for ASR, comparing frame-level teacher-student alignments when the teacher model outputs the word "CAT". Red arrow denotes undesired distillation and blue arrow denotes desired distillation. (a) and (b) include undesired distillation and miss desired distillation while (c) only includes desired distillation.}
    \label{fig:align}
\end{figure*}


Secondly, it is believed that the sparse spike timing of CTC~\cite{graves2006connectionist, senior2015acoustic, chen2016phone} hinders knowledge distillation because it means that the other frames are filled with dummy blank tokens~\cite{senior2015acoustic, takashima2018investigation, kurata2019guiding, tian2022knowledge}. Therefore, selection of informative frames has been regarded as the key to a successful knowledge distillation, and \cite{kurata2019guiding, tian2022knowledge} propose blank frame masking for knowledge distillation by removing blank frames in which the blank token has the maximum probability. However, as shown in Figure~\ref{fig:align}, although blank frame masking is able to alleviate undesired knowledge distillation, it also mask useful frames, that results in a limited performance improvement. Moreover, blank frame is also an important component of the alignment and contributes to the output transcription~\cite{jung23b_interspeech, fan2023key}. For this reason, we claim that the limited performance improvement of knowledge distillation of CTC is not originated from sparse spike timing but the alignment disagreement problem. Thus, we explore the effect of blank frame masking under self-knowledge distillation environment in which the alignment disagreement problem has been mitigated. Unlike the previous teacher-student knowledge distillation method, it is observed that self-knowledge distillation without blank frame masking surpasses the other with blank frame masking, which highlights the importance of distilling blank frames. In other words, we reveal that sparse spike timing is not a major factor of limited performance enhancement under the self-knowledge distillation method.

In summary, we introduce a simple and effective frame-level self-knowledge distillation framework for CTC-based ASR models. We address the alignment disagreement problem of the teacher-student knowledge distillation framework by utilizing the self-knowledge distillation framework. At the same time, we lessen the information loss which stems from blank frame masking strategy of the conventional knowledge distillation methods. To verify this, we examine the belief that blank frames deteriorate the knowledge distillation performance and confirm that it is false when the alignment disagreement problem is insignificant such as in self-knowledge distillation. Based on this result, we experimentally prove the usefulness of the proposed method.

\section{Background}
\label{sec:prev}
\subsection{Connectionist Temporal Classification (CTC)}
Given the input sequence $x=[x_1, ..., x_F]$ where $x_f\in\mathbb{R}^D$ denotes $D$-dimensional acoustic speech feature of frame $f$ and label sequence $y=[y_1, ..., y_N]$ of $y_n\in\mathcal{Y}$ where $\mathcal{Y}$ denotes a label set, automatic speech recognition aims to translate $x$ into $y$. In many cases, however, the sequence length of the input and the label are different (i.e., $F>>N$), and the optimal alignment between them is unknown. CTC~\cite{graves2006connectionist} is a useful framework to address this issue. For input $x$, CTC computes the possible alignment set $y^a=[y^a_1, ..., y^a_F]\in\beta^{-1}(y)$ of $y$ where $y^a_f\in\mathcal{Y'}$ is the element of the blank-augmented label set $\mathcal{Y}'=\mathcal{Y}\bigcup\phi$ and $\phi$ is a blank token. An automatic speech recognition model $h$ with $L$-layer is trained to minimize $\mathcal{L}_\text{CTC}$ under the CTC framework as follows:
\begin{equation}
    \mathcal{L}_\text{CTC}=-\log\sum_{y^a\in\beta^{-1}(y)}p(y^a|x^L),
\end{equation}
where $x^L$ is a $L$-th layer output (i.e., $x^L=[x^L_1,...,x^L_F]=h(x)=h^L\circ h^{L-1}\cdots \circ h^1(x)$).

\subsection{Frame-Level Knowledge Distillation}
\cite{takashima2018investigation} explores the frame-level knowledge distillation method for CTC following the conventional knowledge distillation framework. Frame-level knowledge distillation uses the frame-level output of the teacher model to teach the student model. Formally, the loss of frame-level knowledge distillation with cross-entropy is formulated as follows:
\begin{equation}
    \mathcal{L}^{\text{frame}}_{\text{KD}}=-\sum_{f=1}^F\sum_{a\in\mathcal{Y}'} p(a|x^T_f)\log p(a|x^S_f),
\end{equation}
where $x^T=[x^T_1, ..., x^T_F]$ is the output of the teacher model and $x^S=[x^S_1, ..., x^S_F]$ is the output of the student model.

To stabilize the frame-level knowledge distillation, softmax-level knowledge distillation (Sftmx-KD)~\cite{yoon2021tutornet} is proposed. Softmax-level knowledge distillation substitutes the loss function of the frame-level knowledge distillation with $l_2$ loss function, which is stable due to the $l_2$ loss function being bounded. The loss is formulated as follows:
\begin{equation}
    \mathcal{L}^{\text{sfmx}}_{\text{KD}}=\sum^F_{f=1}\sum_{a\in\mathcal{Y}'}(p(a|x^T_f)-p(a|x^S_f))^2.
\end{equation}

\subsection{Frame-Level Knowledge Distillation with Masking}
Previous studies~\cite{takashima2018investigation, takashima2019investigation} confirm that frame-level knowledge distillation has limited performance improvement. One of the possible reasons is that spike timings are sparse, meaning that considerable frames are filled with a dummy blank token. Guide-CTC (G-CTC)~\cite{kurata2019guiding} challenges this problem by masking the blank frames of the teacher model in knowledge distillation as follows:
 \begin{gather}
  \mathcal{L}^{\text{Guide\text{-}CTC}}_{\text{KD}}=-\sum^F_{f=1}\sum_{a\in\mathcal{Y}'}M_{f,a}\log p(a|x^S_f),\\
  M_{f, a}=\mathds{1}(\arg\max_{\bar{a}\in\mathcal{Y}'}p(\bar{a}|x^T_f)=a\neq\phi),
\end{gather}
where $M_{f, a}$ denotes a mask function. In short, Guide-CTC utilizes the frame-level greedy prediction of the teacher as a label only when the teacher's prediction is not a blank token.


\begin{figure*}
    \centering
    \includegraphics[width=.9\textwidth]{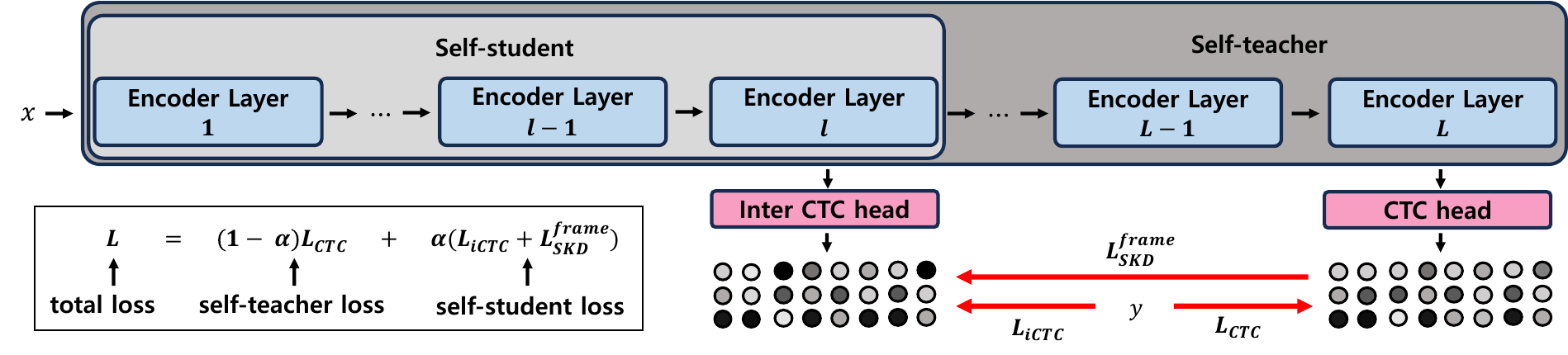}
    \caption{A framework for self-knowledge distillation of $L$-layer CTC-based ASR model with $l$-layer student model. The total training loss contains self-teacher loss and self-student loss.}
    \label{fig:skd}
\end{figure*}

\subsection{Intermediate CTC and Layer Pruning}
Based on the conventional CTC loss, Intermediate CTC~\cite{lee2021intermediate} also utilizes the intermediate output of the $l$-th layer $x^l=[x^l_1, ..., x^l_F]=h^l\circ h^{l-1}\cdots \circ h^1(x)$ to regularize the model using the intermediate CTC loss as follows:
\begin{equation}
    \mathcal{L}_\text{iCTC}=-\log\sum_{y^a\in\beta^{-1}(y)}p(y^a|x^l),
\end{equation}
where $0<l<L$. The total loss function $\mathcal{L}$ with intermediate CTC is formulated as $\mathcal{L}=(1-\alpha)\mathcal{L}_\text{CTC}+\alpha\mathcal{L}_\text{iCTC}$. Upon this, it is possible to directly distill the teacher's output to the student's intermediate outputs as presented in~\cite{yoon2023inter}. Also, \cite{lee2021layer} extends intermediate CTC to a layer pruning method by removing the layers after the intermediate CTC layer. For the experiment, we utilize this method for the layer pruning baseline.

\section{Method}
\subsection{Frame-Level Self-knowledge Distillation}
Despite the success of the teacher-student knowledge distillation framework, knowledge distillation for CTC model suffers from the alignment disagreement problem. The alignments of teacher and student models can be diverse since CTC takes into account every possible alignment between an input and a label for training, meaning that the alignment is not fixed. We tackle this issue by introducing frame-level self-knowledge distillation (SKD) for CTC. The overall architecture of SKD is illustrated in Figure~\ref{fig:skd}. The proposed method utilizes the intermediate CTC framework into the self-knowledge distillation framework. Specifically, the self-teacher model consists of CTC head and layers $1, ..., L$, and the self-student model consists of intermediate CTC head and layers $1, ..., l$ so that $1\leq l<L$ which are shared with the self-teacher. In other words, SKD contains the teacher and student models in a single architecture, and the teacher inherits frame-level information of the student. Thus, the alignment disagreement problem can be intrinsically decreased. Also, SKD is efficient in memory and training time compared to KD because SKD only needs the single architecture. The loss function of the proposed method is formulated as follows:
\begin{equation}
  \mathcal{L}_{\text{SKD}}^\text{frame}=-\sum^F_{f=1}\sum_{a\in\mathcal{Y}'}p(a|x^L_f)\log p(a|x^l_f),
\end{equation}
where a stop gradient operation is applied to the output of the teacher model. Blank frame masking is not applied to the proposed method because we observe that blank frame masking is inappropriate for self-knowledge distillation in which the alignment disagreement is not significant. Detailed experimental results are introduced in section~\ref{sec:results}.
Finally, the total loss function is formulated as follows:
\begin{equation}
    \mathcal{L}=(1-\alpha)\mathcal{L}_{\text{CTC}}+\alpha(L_{\text{iCTC}}+\mathcal{L}^{\text{frame}}_{\text{SKD}}),
\end{equation}
where $\alpha\in[0,1]$ is a hyperparameter that decides the training strength of the student.


\subsection{Frame-Level Self-knowledge Distillation with Scheduling}
The teacher and the student are independent and trained in a row at the conventional teacher-student KD methods. SKD can also directly utilize this strategy by sequentially setting $\alpha=0$ and $\alpha=1$ for the total loss, $\mathcal{L}$. However, setting $\alpha=1$ for self-knowledge distillation leads to a catastrophic forgetting of the self-teacher model since the shared model is trained only with the self-student model's loss without considering the self-teacher model's loss. So, we introduce a scheduling function $\alpha=\tau(e, E)$ such that $\tau(e, E)<1$ for self-knowledge distillation where $e$ is training epoch and $E$ is the total training epoch. Also, we consider $\tau(e, E)$ which satisfies $\frac{1}{E}\sum_{e=1}^E\tau(e, E)=0.5$, meaning that the average of hyperparameter $\alpha$ is fixed to $0.5$ to preserve the total training volume of the student model. Based on the linear scheduling $\tau(e, E)=\dfrac{e-1}{E-1}$ from $\alpha=1$ to $\alpha=0$, we introduce the clipped linear scheduling for $\alpha$ to enforce $\tau(e, E)<1$ by $\tau(e, E)=\min\{\max\{\dfrac{e-1}{E-1}, t\}, 1-t\}$. Following the original intermediate CTC~\cite{lee2021intermediate} which uses $a=0.3$, we empirically set $t=0.3$ for all experiments.

\section{Experiments}
\label{sec:experiment}
\subsection{Experimental Setup}
For the experiment, pre-trained 12-layer transformer encoder-based models are considered: HuBERT Base~\cite{hsu2021hubert} and WavLM Base+~\cite{chen2022wavlm}. $6,8$ and $10$-layer models are used as the student models with the 12-layer teacher models. All models smaller than the 12 layers is simply initialized using the first $l$ layers from the pre-trained models for a fair comparison of the baseline, KD, and SKD methods. We also experiment with a larger model setting using an 18-layer teacher model and a 12-layer student model. To initialize the 18-layer teacher model, the last 6 layers from the 12-layer pre-trained models are duplicated and repeated after the 12 layers.

\begin{table*}[ht!]
\centering
\caption{Overall performances in terms of WER (\%) of the LibriSpeech dataset. Four different student layers are utilized: 12, 10, 8, 6.}
\begin{tabular}{clcccccccc}
\toprule
    \multicolumn{1}{c}{\multirow{3}{*}{Model}} & \multicolumn{1}{c}{\multirow{3}{*}{Method}} & \multicolumn{2}{c}{$18\rightarrow 12$} & \multicolumn{2}{c}{$12\rightarrow 10$} & \multicolumn{2}{c}{$12\rightarrow 8$} & \multicolumn{2}{c}{$12\rightarrow 6$}\\ \cmidrule(lr){3-4}\cmidrule(lr){5-6}\cmidrule(lr){7-8}\cmidrule(lr){9-10}
    & & test-clean & test-other & test-clean & test-other & test-clean & test-other & test-clean & test-other\\\midrule
    \multirow{6}{*}{HuBERT} & Baseline (T) &  5.8 & 13.0 & 6.3 & 14.4 & 6.3 & 14.4 & 6.3 & 14.4\\
    & Baseline (S) & 6.3 & 14.4 & 10.3 & 22.8 & 18.4 & 36.2 & 34.9 & 55.8\\\cmidrule(lr){2-10}
    & G-CTC (S) & 6.0 & 13.7 & 8.7 & 19.7 & 16.8 & 33.8 & 31.7 & 53.1\\
    & Sftmx-KD (S) & 6.4 & 14.6 & 9.8 & 22.0 & 21.8 & 41.0 & 34.7 & 55.7\\
    & Layer Prun (S) & \textbf{5.6} & \textbf{13.3} & 8.4 & 20.2 & 16.3 & 34.2 & 30.9 & 53.1\\
    & SKD (S) & 5.8 & 13.5 & \textbf{8.1} & \textbf{19.4} & \textbf{15.5} & \textbf{33.4} & \textbf{28.3} & \textbf{51.1}\\\midrule
    \multirow{6}{*}{WavLM} & Baseline (T) &  4.7 & 9.8 & 5.2 & 11.4 & 5.2 & 11.4 & 5.2 & 11.4\\
    & Baseline (S) & 5.2 & 11.4 & 9.6 & 21.5 & 20.7 & 38.8 & 28.9 & 51.1\\\cmidrule(lr){2-10}
    & G-CTC (S) & 8.5 & 12.5 & 9.3 & 18.0 & 18.5 & 34.1 & 26.9 & 48.3\\
    & Sfmx-KD (S) & 5.2 & 11.6 & 9.6 & 21.5 & 18.5 & 37.1 & 28.2 & 50.7\\
    & Layer Prun (S) & 7.2 & 11.5 & 9.2 & 17.1 & 14.3 & 30.8 & 25.8 & 48.6\\
    & SKD (S) & \textbf{5.0} & \textbf{11.0} & \textbf{ 7.1} & \textbf{16.8} & \textbf{13.4} & \textbf{30.0} & \textbf{22.8} & \textbf{46.0}\\
    \bottomrule
\end{tabular}
\label{tab:main}
\end{table*}

\begin{table}[ht!]
\centering
\caption{WER (\%) of LibriSpeech test-other dataset. Student models are distilled from the 12-layer HuBERT Base model.}
\begin{tabular}{clcc}
\toprule
    \multicolumn{1}{c}{\multirow{3}{*}{\# student layers}} & \multicolumn{1}{c}{\multirow{3}{*}{Method}} & \multicolumn{2}{c}{test-other WER (\%) }\\ \cmidrule{3-4}
    & & \textit{w/ mask}& \textit{w/o mask}\\ \midrule
    \multirow{2}{*}{10 layers} & G-CTC & 19.7 & 19.7  \\
    & SKD & 19.9 & \textbf{19.4 } \\\midrule
    \multirow{2}{*}{8 layers} & G-CTC & 33.8 & 34.3 \\
    & SKD & 33.3 & \textbf{33.0}\\ \midrule
    \multirow{2}{*}{6 layers} & G-CTC & 53.1 & 53.5 \\
    & SKD & 51.7 & \textbf{51.1} \\
    \bottomrule
\end{tabular}
\label{tab:mask}
\end{table}

For fine-tuning, we use 100-hour train-clean subset from LibriSpeech~\cite{panayotov2015librispeech} dataset. We use the learning rate of 3e-5 with AdamW~\cite{loshchilov2018decoupled} optimizer using 10\% warmup steps and the cosine learning rate decaying for $E=200$. The models are optimized with a batch size of 128. The feature encoder is fixed during training and the model is freezed for the first 12.5\% of the steps except the linear CTC head. We use a character-level tokenizer, and the language model is not utilized for decoding. Other details follow the original HuBERT and WavLM models. Note that the baseline results are reproduced.

\subsection{Results}
\label{sec:results}
Table~\ref{tab:main} shows the word error rate (WER, \%) of the automatic speech recognition task. We compare the two KD methods including Guide-CTC (G-CTC) and softmax-based KD (Sfmx-KD), one layer pruning method using an intermediate CTC (Layer Prun), and the proposed self-knowledge distillation method (SKD). It is observed that the KD-based methods and the layer pruning method improve performance over the baselines for most of the cases. However, the proposed SKD method outperforms the other methods except for a single case: the 18-layer HuBERT teacher with the 12-layer student. For instance, the WavLM 8-layer model achieves 20.7\% WER for test-clean dataset, and the layer pruning achieves 14.3\%, surpassing the 18.5\% of G-CTC and Sfmx-KD. SKD achieves 13.4\%, which is the lowest WER.

We analyze the effect of blank frame masking which have been regarded as an important component of KD. Table~\ref{tab:mask} shows that the conventional KD (Guide-CTC) with blank frame masking contributes to a performance improvement compared to the Guide-CTC without the masking function. However, we observe that distilling the entire frame without masking for SKD leads to an even lower WER. This supports our argument that distilling blank frames also contributes to an increased performance when the alignment disagreement of the teacher and student is not significant, as depicted in Figure~\ref{fig:align} (c) for which the alignment disagreement is not as severe as in Figure~\ref{fig:align} (a). Furthermore, we examine the frame-level teacher-student output accuracy with and without the masking, as shown in Table~\ref{tab:spikeacc}. Overall, SKD achieves a similar or higher frame-level output accuracy between the teacher model and the student model than Guide-CTC, which accounts for the improved performance of SKD. Guide-CTC with masking shows the worst total frame accuracy (total) for in all experiments. Guide-CTC without masking increases total accuracy but decreases the non-blank frames' teacher-student output accuracy (active). Still, SKD without masking achieves higher active accuracy than Guide-CTC with masking while achieving the highest total accuracy except in a single setting.


\begin{table}[t]
\centering
\caption{Accuracy (\%) of frame-level outputs between the teacher model and the student model, measured under two settings: total frames and active frames. The former considers all frames, while the latter excludes blank frames. The experiments are conducted based on HuBERT models.}
\begin{tabular}{clcccc}
\toprule
    \multicolumn{1}{c}{\multirow{3}{*}{\# layers}} & \multicolumn{1}{c}{\multirow{3}{*}{Method}} & \multicolumn{2}{c}{test-clean (\%)} & \multicolumn{2}{c}{test-other (\%)}\\ 
    \cmidrule(lr){3-4}\cmidrule(lr){5-6}
    &  & active & total  & active & total \\ \midrule
    \multirow{4}{*}{$18 \rightarrow 12$} & G-CTC  w/ m &  96.62 & 76.58 & 96.13 & 77.53   \\
    & G-CTC  w/o m & 95.21 & 94.66  & 93.13 & 93.49   \\
    & SKD w/ m & \textbf{99.13} & 88.89  & \textbf{98.58} & 86.15   \\
    & SKD w/o m & 98.83 & \textbf{98.31}  & 98.30 & \textbf{97.77}  \\\midrule
    \multirow{4}{*}{$12 \rightarrow 8$} & G-CTC  w/ m & 92.15 & 78.43 &  86.27  & 76.92    \\
    & G-CTC w/o m & 88.33 & 90.35  & 86.59 & \textbf{90.35}  \\
    & SKD w/ m & \textbf{93.72} & 88.43  & \textbf{89.14} & 85.94  \\
    & SKD w/o m & 91.48 & \textbf{92.69} & 86.47 & 89.62   \\
    \bottomrule
\end{tabular}
\label{tab:spikeacc}
\end{table}

\section{Conclusion}
In this paper, we introduce a self-knowledge distillation method for CTC-based ASR model. Experimental results confirm that the proposed method outperforms the conventional knowledge distillation methods and the layer pruning method across the various model settings. The proposed method is effective in that the proposed method only requires extra linear CTC head computation to the teacher model since the teacher model and the student model share the layers. Furthermore, we investigate the necessity of the blank frame masking for distillation, and observe that the blank frame masking diminishes the performance improvement of self-knowledge distillation. For future work, we consider adapting different teacher-student architectures to the proposed method.

\section{Acknowledgements}
This work was partly supported by Institute of Information \& communications Technology Planning \& Evaluation (IITP) grant funded by the Korea government(MSIT) (No. RS-2022-II220641, XVoice: Multi-Modal Voice Meta Learning, 1/2) and (No. RS-2022-II220320, Artificial intelligence research about cross-modal dialogue modeling for one-on-one multi-modal interactions, 1/2).

\bibliographystyle{IEEEtran}
\bibliography{mybib}

\end{document}